\documentclass[3p,twocolumn]{elsarticle}

\usepackage{graphicx,amsmath, amssymb}
\usepackage{url}

\begin{document}

\begin{frontmatter}

\title{Global Flow of Glasma in High Energy Nuclear Collisions}

\author{Guangyao~Chen}
\author{Rainer~J.~Fries}
\ead{rjfries@comp.tamu.edu}

\address{Cyclotron Institute and Department of Physics and Astronomy,
Texas A\&M University, College Station TX 77843, USA}

\begin{abstract}
We discuss the energy flow of the classical gluon fields created in collisions
of heavy nuclei at collider energies. We show how the Yang-Mills  analoga of
Faraday's Law and Gauss' Law predict the initial gluon flux tubes to
expand or bend. The resulting transverse and longitudinal structure of the
Poynting vector field has a rich phenomenology. Besides the well known radial
and elliptic flow in transverse direction, classical quantum chromodynamics predicts
a rapidity-odd transverse flow that tilts the fireball for non-central
collisions, and it implies a characteristic flow pattern for collisions
of non-symmetric systems $A+B$.
The rapidity-odd transverse flow translates into a directed particle flow $v_1$
which has been observed at RHIC and LHC. The global flow fields in heavy
ion collisions could be a powerful check for the validity of classical
Yang-Mill dynamics in high energy collisions.
\end{abstract}


\end{frontmatter}

\section{Introduction}

Hadrons and nuclei colliding at very large energies can be used to study a
novel regime of quantum chromodynamics (QCD) called the color glass condensate
(CGC) \cite{Gelis:2010nm}. It is characterized by a saturation of the transverse densities
of gluons $\sim 1/Q_s^2$ in the initial wave functions, setting an
energy scale $Q_s$, the saturation scale. High occupation numbers permit  a
quasi-classical description of the gluon fields in the nuclei before collisions
\cite{McLerran:1993ka,McLerran:1993ni}, and at very early
time after the collision \cite{KoMLeWei:95}.
Data from the Relativistic Heavy Ion Collider (RHIC) and the Large Hadron
Collider (LHC) have given us hints that we have reached an energy regime
where CGC could be the correct description of the relevant degrees of
freedom of early stages of nuclear collisions.

There is a large body of evidence that a thermalized quark gluon
plasma (QGP) is eventually created within the first fm/$c$ after the collision
of nuclei at RHIC or LHC energies \cite{Adcox:2004mh,Adams:2005dq}.
CGC as an effective theory can be used to describe the very first step in this
process, from the original nuclear wave functions to a
space-time volume filled with coherent and far off-equilibrium gluon fields,
called a glasma,
at a time $\tau_0 \sim 1/Q_s \sim 0.1 \ldots 0.2$ fm/$c$ after the
collision. The dominant feature of the fields immediately after the collision
are flux tubes of longitudinal chromo-electric and
chromo-magnetic fields \cite{Fries:2005yc,Fries:2006pv,Lappi:2006fp}.
After this initial period the fields have to decohere and
the emerging particles have to approach equlibrium 
\cite{Fries:2008vp,Romatschke:2006nk,Gelis:2007pw}.
After equilibrium is reached at a time $\tau_{\mathrm{th}} > \tau_0$
hydrodynamics can be used to describe the global collective motion of the
system \cite{Kolb:2003dz,Song:2007ux,Schenke:2010nt,Fries:2010ht}.
The success of hydrodynamics in heavy ion collisions is based on its ability to
capture key features of the collective flow of the thermalized QGP, foremost
the radial transverse flow $v_r$ and the quadrupole asymmetry $v_2$,
called elliptic flow.

However, flow is not a privilege of thermalized matter with
pressure gradients. In fact the effective ``transverse pressure'' $p_T = T^{11}=T^{22}$
in a purely longitudinal classical gauge field is equal to the
energy density $T^{00}$, and thus gradients in pressure would be larger
by a factor 3 than in a relativistic free gas with the same energy density
\cite{Fries:2005yc,Vredevoogd:2008id}. Thus the amount of initial collective
motion, often termed pre-equilibrium flow, at the starting time $\tau_{\mathrm{th}}$ of the
hydrodynamic evolution is not expected to be small. In fact various efforts have been made to 
estimate the magnitude of transverse flow around midrapidity
from color glass  or other dynamics \cite{Vredevoogd:2008id,Schenke:2012wb},
although many hydrodynamics simulations still neglect pre-equilibrium flow.

In this Letter we expand existing calculations by analyzing the full
3+1-dimensional structure of the initial gluon energy flow.
To that end we will present an analytic calculation of  the chromo-electric and
chromo-magnetic fields $\vec{E}$ and $\vec{B}$ immediately after the
collision of nuclei in a strictly classical implementation of CGC, the
McLerran-Venugopalan (MV) model \cite{McLerran:1993ka,McLerran:1993ni}.
We then compute the energy momentum tensor
$T^{\mu\nu}$ of the gluon field as a function of space-time coordinates for
small proper times $\tau=\sqrt{t^2-z^2}$ after the collision. We analyze the
resulting global map of energy and momentum flow, in particular the transverse
Poynting vector $S^i=T^{0i}$, $i=1,2$. We will argue that it can be
decomposed into two parts, $S_+^i$ and $S_-^i$.  $S_+^i$ is even in
the space-time rapidity  $\eta=1/2 \ln\left[(t+z)/(t-z)\right]$ and
driven by the gradient of the initial transverse pressure $p_T$ of the gluon
field; in short this term behaves ``hydro-like'' and has in fact been
discussed before analytically and numerically
\cite{Fries:2005yc,Vredevoogd:2008id,Schenke:2012wb}.
The novel $\eta$-odd term $S_-^i$ arises from asymmetries between the
two nuclei, e.g.\ for collisions $A+B$ of different nuclei or for symmetric
$A+A$ collision with finite impact parameter $b$.
Since it vanishes at mid-rapidity it has not been discussed in situations where
one expects flow to be (almost) boost-invariant.
One can show that despite the superficial appearance the $\eta$-odd flow term
$S_-^i$ does not break the boost-invariance which is intrinsic to the MV model.
We will argue, using the abelian case, that both rapidity-odd and -even terms
are a natural consequence of Faraday's Law and Gauss' Law.
We will conclude this Letter by discussing the phenomenological consequences
for several colliding systems and impact parameters.

\section{Gluon Fields in the MV Model}

In the CGC the large momentum components of the nuclear wave function are assumed
to be strongly Lorentz-contracted in the lab frame and can be approximated by an
infinitely thin sheet.  The large momentum quarks and gluons form an
effective color $SU(3)$ charge density $\rho_{\underline{a}}(\vec r)$,
$\underline{a} = 1,\ldots N_c^2-1$, on that sheet,
where $\vec r$ is a transverse position. They act as sources for
the soft gluon modes which are described by the classical field $A^\mu$.
In the McLerran-Venugopalan model the source $\rho_{\underline{a}}$ moving
along the $+$-light cone leads to a $SU(3)$-current $J^\mu = \delta^{\mu+} \delta(x^-)
\rho_{\underline{a}}  t^{\underline{a}}$
in light cone gauge. The soft gluon modes are determined by the Yang-Mills
equation $D_\mu F^{\mu\nu} = J^\nu$.
The charge density $\rho_{\underline{a}}(\vec r)$ varies on time scales that are large
compared to typical time scales of the collision. It vanishes on average,
$\langle \rho \rangle = 0$, due to local color neutrality, but has a
non-vanishing variance $\mu$ which in the MV model is realized by a
Gaussian distribution with
\begin{equation}
  \label{eq:chargedensnorm}
  \langle \rho_{\underline{a}}(\vec r_1) \rho_{\underline{b}}
  (\vec r_2) \rangle = \frac{g^2}{N_c^2-1} \delta_{\underline{ab}}
  \mu(\vec r_1) \delta^2 (\vec r_1 - \vec r_2 )  \, .
\end{equation}
Thus $\mu$ is  the average, \emph{color-summed}, squared density of charges in the
transverse plane (which differs from the definition in \cite{Lappi:06} by a factor $N_c^2-1$).

The most fundamental result of the MV model is an expression for the
gluon distribution function of a nucleus in light cone gauge ($i,j=1,2$)
\begin{equation}
  \label{eq:gluondist}
  \langle A^i_{\underline{a}}(\vec r) A^j_{\underline{b}} (\vec r)\rangle =  
  \frac{g^2}{8\pi (N_c^2-1)}  \delta_{\underline{ab}}\delta^{ij} \mu(\vec r) \ln \frac{Q^2}{m^2}
\end{equation}
where the UV limit (both fields evaluated at the same transverse position
$\vec r$) has been taken. Here $Q$ is a UV cutoff related to the resolution with which the
gluon distribution is probed and $m$ is an infrared scale that can be
introduced as an effective gluon mass. This result was first calculated in
Ref.\ \cite{JMKMW:96} for infinitely large, homogeneous nuclei, i.e.\ for constant
source area density $\mu$. Of course this is not a realistic approximation for
nuclei. We have checked that if variations of  $\mu(\vec r)$ are permitted but
are small on the IR length scale $\sim 1/m$, i.e.\
\begin{equation}
  \label{eq:conditionv}
  m^{-1} |\nabla \mu(\vec r)| \ll \mu(\vec r)
\end{equation}
many well-known results of the MV model, in particular
Eq.\ (\ref{eq:gluondist}) hold and corrections can be classified in an
expansion in gradients $m^{-1}\nabla^i \mu(\vec r)$. In simple terms this
indicates that the physics of color glass and the less known QCD dynamics at
larger distances can be separated in a meaningful way as long
as condition (\ref{eq:conditionv}) is satisfied. We will report details of
this calculation elsewhere \cite{CFKL:13}.

Given the purely transverse gauge fields $A_1^i=A^i_{\underline{a},1}(\vec r)
t^{\underline{a}}$ and $A_2^i=A^i_{\underline{a},2}(\vec r) t^{\underline{a}}$
of the nuclei before the collision, in their respective light cone gauge, one
can write the longitudinal chromo-electric and -magnetic fields after the collision for
small times $\tau$ as \cite{Fries:2006pv,CFKL:13,Fujii:2008km}
\begin{align}
 E^3 &= \left(1 + \frac{\tau^2}{4} D^2 \right) E_0
 + \mathcal{O}(\tau^4)  \\
 B^3 &= \left(1 + \frac{\tau^2}{4} D^2 \right) B_0 + \mathcal{O}(\tau^4)
\end{align}
where $D^2 =D^iD^i$ is the square of the covariant derivative with respect to gauge field
$A_1^i+A_2^i$ and $E_0 =  ig\left[ A_1^j, A_2^j \right] $, $B_0 = ig \epsilon^{jk} \left[ A_1^j,
 A_2^k \right] $ are the longitudinal fields at $\tau=0$.
Since the transverse fields vanish at $\tau=0$ this has given rise to the
notion of flux tubes of longitudinal electric and magnetic fields
\cite{Lappi:2006fp}. In terms of those initial longitudinal fields the
transverse fields can be expressed as
\cite{Fries:2006pv,CFKL:13,Fujii:2008km}
\begin{align}
  \label{eq:ei}
  E^i & = -\frac{\tau}{2}\left(\sinh \eta\, D^i E_0 + \cosh\eta\, \epsilon^{ij} D^j
  B_0 \right)    \\ 
  \label{eq:bi}
  B^i & = -\frac{\tau}{2}\left( \sinh \, \eta D^i B_0 - \cosh\eta\, \epsilon^{ij}
  D^j E_0  \right)  
\end{align}
plus terms of order $\tau^3$ and higher.

It is straightforward to write down the initial energy density $\epsilon_0 =T^{00}(\tau=0)=
(E_0^2+B_0^2)/2$ and the initial transverse energy flux given by the Poynting
vector $T^{0i}$. The leading behavior at small $\tau$ has four terms, two even in $\eta$,
\begin{equation}
  \label{eq:2a}
  S^i_+ = \frac{\tau}{2} \cosh\eta \left(E_0 D^i E_0 + B_0 D^i B_0
  \right)  \, ,
\end{equation}
and two odd in $\eta$,
\begin{equation}
  \label{eq:2b}
  S^i_- = \frac{\tau}{2} \sinh\eta \, \epsilon^{ij} \left(E_0 D^j B_0 - B_0 D^j E_0
  \right)  \, ,
\end{equation}
and $T^{0i}=S^i_+ + S^i_-$.
The longitudinal energy flow is suppressed and starts at order
$\mathcal{O}(\tau^2)$,
\begin{multline}
  T^{03} = \frac{\tau^2}{8} \left[ \sinh 2\eta \left( (DE_0)^2 + (DB_0)^2
    \right) \right. \\ + \left.
  2 \cosh 2\eta  \,\epsilon^{ij} (D^i E_0)(D^j B_0) \right] \, .
\end{multline}
Note that the MV model is boost-invariant by definition and the
$\eta$-dependence of the energy flow does not violate boost-invariance.
One can verify that the full energy momentum tensor $T^{\mu\nu}$
satisfies $\Lambda^{\mu\mu'}(y) T_{\mu'\nu'}(\eta) \Lambda^{\nu\nu'}(y) =
T_{\mu\nu}(\eta+y)$ where $\Lambda(y)$ is the Lorenz boost tensor with
rapidity $y$ along the $z$-axis.
In fact the rapidity-odd field is a natural consequence of the field equations
as the next section will show.

\section{An Electrodynamic Analogon}

\begin{figure}[tb]
  \includegraphics[width=\columnwidth]{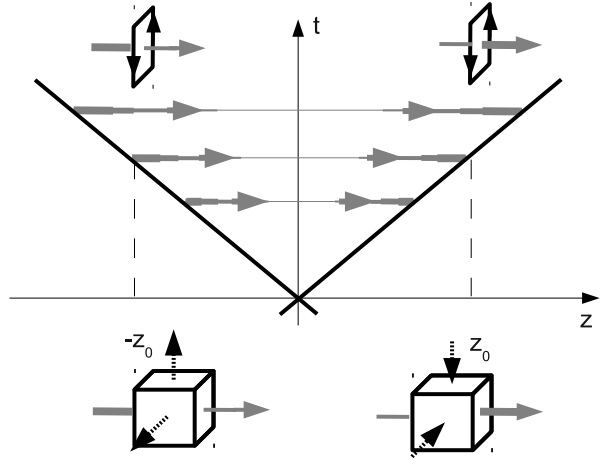}
  \caption{Two observers at $z=z_0$ and $z=-z_0$ test Amp\`ere's and Faraday's
  Laws with areas $a^2$ in the transverse plane and Gauss' Law with a cube of
  volume $a^3$. The transverse fields from Amp\`ere's and Faraday's Laws
  (black solid arrows) are the same in both cases, while the transverse fields from 
  Gauss' Law (black dashed arrows) are
  observed with opposite signs. Initial longitudinal fields are indicated by
  solid grey arrows, thickness reflects field strength.}
  \label{fig:2}
\end{figure}

Let us consider the following equivalent boundary value
problem in classical electrodynamics.  In the forward light cone $\tau >0 $
we have the Maxwell Equations
  $\partial_\mu F^{\mu\nu} = 0$
without sources.  On the light cone $\tau=0$ we demand the boundary conditions
$\vec E(\tau =0,\vec r) = E_0 (\vec r)\vec e_z$,  $\vec B(\tau
=0,\vec r)  = B_0 (\vec r)\vec e_z$, i.e. the initial fields are
purely longitudinal. We also assume that those fields are related
through transverse fields $A_1^i$ and $A_2^i$ as $E_0 = \delta^{ij}  A_1^i
A_2^i$ and $B_0 = \epsilon^{ij}  A_1^i A_2^j$.
The abelian problem for fixed initial conditions can in principle be solved
analytically \cite{KoMLeWei:95,CFKL:13}, but it will suffice here to give the
solution order by order
in powers of $\tau$ as we did in the case of QCD.
From the QCD solutions we can immediately conclude that
the longitudinal fields in the abelian case are
\begin{align}
   E^3 &= \left(1 + \frac{t^2-z^2}{4} \nabla^2 \right) E_0
   \\
   B^3 &= \left(1 + \frac{t^2-z^2}{4} \nabla^2 \right) B_0  \, ,
\end{align}
while the transverse fields are
\begin{align}
  E^i & = \frac{z}{2} \nabla^i E_0 + \frac{t}{2} \epsilon^{ij} \nabla^j
  B_0  \\
  B^i & = \frac{z}{2} \nabla^i B_0 - \frac{t}{2} \epsilon^{ij}
  \nabla^j E_0   \, ,
\end{align}
for small times $\tau$, i.e.\  $t^2 \approx z^2$.
The cartesian coordinates permit simple checks of these solutions with
Gauss', Amp\`ere's and Faraday's Laws.

\begin{figure}[tb]
  \includegraphics[width= \columnwidth]{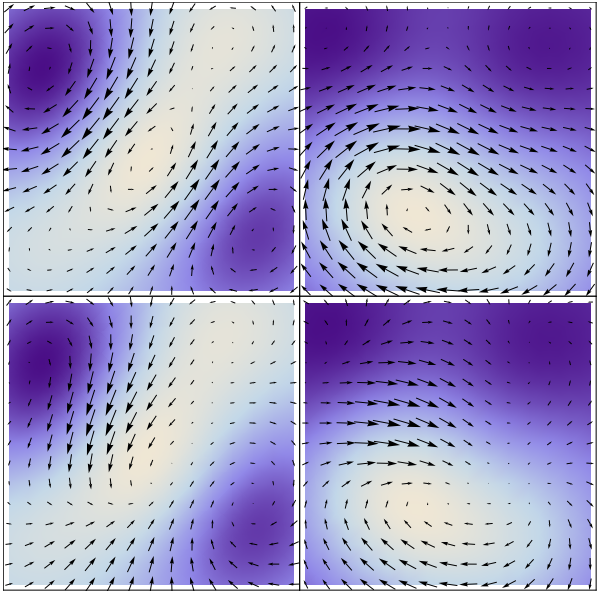}
  \caption{Transverse electric fields (left panels) and
    magnetic fields (right panels) at $\eta=0$ (upper panels) and $\eta=1$
    (lower panels) in an abelian example for a random distribution of
    fields $A^i_1$, $A^i_2$. The initial longitudinal fields
    $B_0$ (left panels) and $E_0$ (right panels) are indicated through the
    density of the background (lighter color = larger values). At $\eta=0$ the
    fields are divergence-free and clearly following Amp\'ere's and Faraday's Laws, respectively.}
  \label{fig:3}
\end{figure}

\begin{figure}[tb]
  \includegraphics[width= \columnwidth]{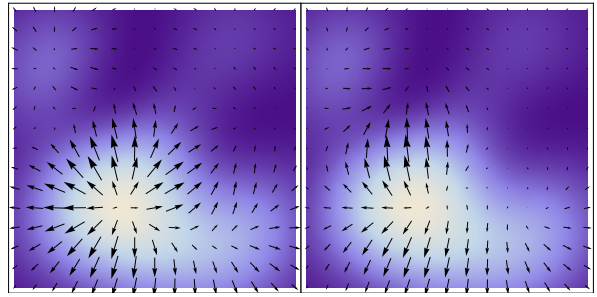}
  \caption{Example for transverse flow of energy for $\eta=0$
    (left panel) and $\eta=1$ (right panel) in the abelian example for
    the same random distribution of fields $A_1^i$, $A_2^i$ as in
    Fig.\ \ref{fig:3}. The initial energy density $T^{00}$ is shown through the
    density of the background (lighter color = larger values). At
    $\eta=0$ the flow follows the gradient in the energy density in a
    hydro-like way while away from mid-rapidity energy flow gets quenched in 
    some directions and amplified in others.}
  \label{fig:4}
\end{figure}

There is a straight-forward interpretation of some aspects of these
results. Let us choose, just as an example, a transverse position where
$E_0, B_0 > 0$ and $\nabla^2 E_0$, $\nabla^2 B_0 < 0$ so that the longitudinal
fields decrease away from the light cone $t^2=z^2$. Two observers at fixed points
$z=z_0 > 0$ and $z=-z_0$ would observe the same electric (magnetic) flux through a small
transverse area $a^2$ with an initial value $E_0a^2$ ($B_0a^2$) at $t=z_0$
which then diminishes at the same rate $\nabla^2 E_0 a^2t/2$ ($\nabla^2 B_0
a^2t/2$) for both.  Due to Amp\`ere's (Faraday's) Law this reduction induces
magnetic (electric) fields curling with a negative (positive) chirality around
the longitudinal fields, respectively, see Fig.\ \ref{fig:2}.

On the other hand the same two observers at fixed points $z_0$ and
$-z_0$ can at time $t=z_0$ count the electric or magnetic flux through small cubes of
volume $a^3$ whose  sides are aligned with the coordinate axes. One side is held at
$z=\pm z_0$, while the opposite side is at $z=\pm z_0 \mp a$ for the
observer at $z_0$ or $-z_0$ respectively. In the former case the total flux
out of the box due to the longitudinal field is $-z_0a^3\nabla^2 E_0/2 > 0$
while for the observer at $-z_0$ the net flux of longitudinal field has the
opposite sign. Thus at $z_0$
a net flux of transverse field has to enter into the box while at $-z_0$ the
same amount has to flow out of the box to satisfy Gauss' Law.

To summarize, the transverse fields naturally have a part due to Gauss' Law
with vanishing circulation ($\epsilon^{ij} \nabla^j$), which is odd in $\eta$,
and they have a part due to Amp\`ere's and Faraday's Law (and with different
signs between the magnetic and electric part due to the Lenz rule)
with vanishing transverse divergence ($\nabla^i$),  which is even in $\eta$.
Fig.\ \ref{fig:3}  shows the transverse electric and magnetic fields for two rapidities
$\eta$ for random fields $A_1^i$ and $A_2^i$ in a sector of the transverse plane.
One can check that this statement about transverse fields translates directly
into a matching statement about the transverse flow of energy since the
initial transverse Poynting vector is linear in the transverse fields,
$T^{0i} = \epsilon^{ij} (E^j B_0 - B^j E_0)$. Thus we have
the four contributions already discussed in the case of QCD, two of them odd in $\eta$.
Fig.\ \ref{fig:4} shows the flux of energy in the transverse plane for two
rapidities for the same random configuration of fields $A_1^i$, $A_2^i$.

\section{Averaging Over Field Configurations}

Eqs.\ (\ref{eq:2a}), (\ref{eq:2b}) can be the basis for a Monte Carlo
modeling of the energy flow starting from a sampling of possible nuclear
fields $A_1^i$, $A_2^i$ or initial charge densities $\rho_1^{\underline{a}}$,
$\rho_2^{\underline{a}}$ in the colliding nuclei. Here we will present a
calculation of the expectation value of the flow fields carried out in the
MV model with slowly varying average charge density $\mu$.
It has been shown in \cite{Lappi:06} that the expectation value of the initial
energy density is given by the product of the average source densities
$\mu_1$, $\mu_2$ in both nuclei,
\begin{equation}
 \varepsilon_0 (\vec r) =
 \frac{g^6}{32\pi^2}\frac{N_c}{N_c^2-1} \ln^2 \frac{Q^2}{m^2} \mu_1 (\vec r)
 \mu_2(\vec r)   \, .
\end{equation}

For the transverse Poynting vector let us discuss our expectations before
calculating the result. First we note that the non-abelian terms from
the covariant derivatives would lead to expectation values of three-gluon
correlators $\langle AAA\rangle$ in one of the nuclei which under the
assumptions of the McLerran-Venugopalan model vanish. Furthermore we can
convince ourselves that because of the relation between $E_0$ and $B_0$ given
by $A_1^i$ and $A_2^i$ the two terms in (\ref{eq:2b}) are the same up to a
sign and thus add up constructively.
One can also easily argue that for dimensional reasons
$\mu_1\nabla^i \mu_2 + \mu_2\nabla^i \mu_1$ and
$\mu_1\nabla^i \mu_2 - \mu_2\nabla^i \mu_1$, and their contractions with
$\epsilon^{ij}$ are the only four transverse vectors available after
averaging.

The first option, symmetric in the two nuclei and circulation-free, is picked
by the $\eta$-even flow term
\begin{multline}
  S^i_+  =  - \frac{\tau}{2}\cosh\eta
  \frac{g^6}{32\pi^2}\frac{N_c}{N_c^2-1} \ln^2 \frac{Q^2}{\hat m^2} 
   \nabla^i\left( \mu_1 \mu_2 \right) \\  = - \frac{\tau}{2}\cosh\eta \nabla^i \epsilon_0
   \, .
\end{multline}
It mimics hydrodynamic flow.
For the $\eta$-odd flow term we need to know the expectation value
of two gluon fields in light cone gauge with one derivative. Following the
same recipe that leads to the expression (\ref{eq:gluondist}) for the gluon
distribution and using our smoothness condition on $\mu(\vec r)$ we obtain
\begin{multline}
  \left\langle \left(\partial^k A^i_{\underline a}\right) (\vec x_\perp ) A^j_{\underline b}(\vec x_\perp)
  \right\rangle =
  \frac{ g^2 }{16\pi(N_c^2-1)}\delta_{\underline{ab}}  \\ \times
  \ln\frac{Q^2}{ m^2} \nabla^l \mu(\vec x_\perp) 
  \left( \delta^{jl} \delta^{ik} - \delta^{il}\delta^{jk} - \delta^{kl}
  \delta^{ij} \right)
\end{multline}
as the leading term contributing to $S^i_-$.
We refer the reader to a detailed calculation of this correlation function in
a forthcoming publication \cite{CFKL:13}.
It is then straight forward to obtain the rapidity-odd flow as
\begin{multline}
  S^i_- = - \frac{\tau}{2}\sinh\eta
  \frac{g^6}{32 \pi^2}\frac{N_c}{N_c^2-1}  \\ \times \ln^2 \frac{Q^2}{m^2}
  \left( \mu_2 \nabla^i \mu_1 - \mu_1\nabla^i \mu_2 \right) \, .
\end{multline}

\section{Discussion of Results}

\begin{figure}[tb]
  \includegraphics[width= \columnwidth]{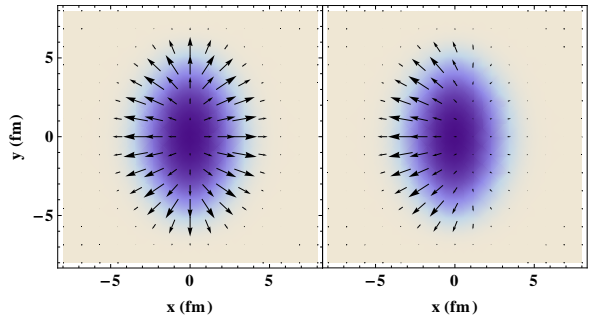}
   \caption{Flow field $V^i$ (black arrows) and energy density $\epsilon_0$
    (shading) in the transverse plane for Au+Au collisions at
    $b=6$ fm. The nucleus centered at $x=3$ fm travels into the plane which is
    the positive $\eta$-direction.
    Left Panel: $\eta=0$. Right Panel: $\eta=1$.}
  \label{fig:5}
\end{figure}

\begin{figure}[tb]
  \begin{center}\includegraphics[width= 4.3 cm]{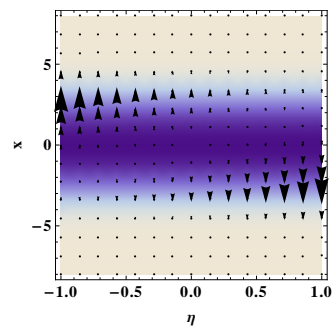}
  \end{center}
  \caption{Same as Fig.\ \ref{fig:5} but plotted in the $\eta-x$-plane defined
    by $y=0$. The flow will lead to a tilted fireball.}
  \label{fig:6}
\end{figure}

\begin{figure}[tb]
  \includegraphics[width= \columnwidth]{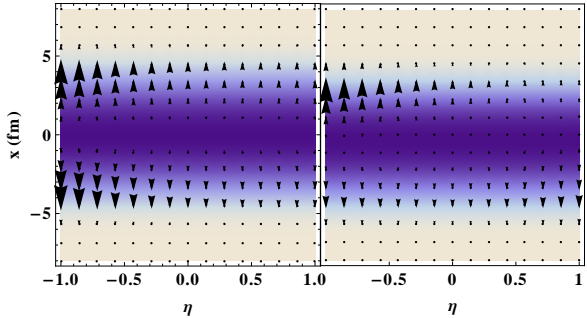}
  \caption{The same as Fig.\ \ref{fig:6} for Au+Cu (Au traveling to
    the right). Left Panel: $b=0$ fm. Right Panel: $b=2$ fm.}
  \label{fig:7}
\end{figure}

\begin{figure}[tb]
  \includegraphics[width= \columnwidth]{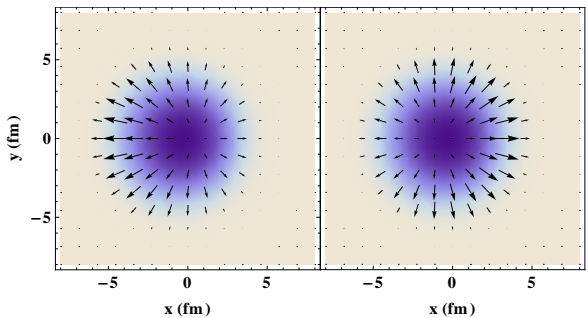}
  \caption{The same as Fig.\ \ref{fig:5} for
    Au+Cu at $b=2$ fm. Left Panel: $\eta=1$, Right Panel: $\eta=-1$.}
  \label{fig:8}
\end{figure}

To summarize, the prediction of classical QCD for the initial average
transverse energy flow
normalized by the average initial energy density is
\begin{multline}
  V^i = \frac{T^{0i}}{\epsilon_0} = - \frac{\tau}{2} \left( \cosh\eta
  \frac{\nabla^i \left(\mu_1\mu_2\right) }{\mu_1\mu_2} \right. \\ \left. + \sinh\eta
  \frac{\mu_2\nabla^i \mu_1 - \mu_1 \nabla^i\mu_2 }{\mu_1\mu_2}
  \right) 
\end{multline}
which is independent of the UV cutoff and the IR regulator.
In the following we have calculated $V^i$ in several situations using
Woods-Saxon profiles for incident nuclei. Fig.\ \ref{fig:5} shows the
average flow field $V^i$ for the collision of two gold nuclei at impact
parameter $b=6$ fm in the transverse plane for two space-time rapidities. 
One can clearly see the evolution in rapidity from a hydro-like flow field at
$\eta=0$ to a preferred flow direction at forward rapidity.

In Fig.\ \ref{fig:6} the same collision is shown in the
$\eta-x$-plane. Clearly the flow tilts the fireball clockwise. The orientation
of rotation is as if the gluon flux tubes preferred to expand in the wake
of spectator nucleons in such a way that the flow increases with increasing
separation from the spectators in rapidity. 
However this can not be taken literally as the origin of
the effect. Our calculation is based on a small time expansion and
the response of the energy density to the flow will come in at the next order.
Note that the normalizations of the vector fields in the figures are arbitrary.
Typical values of the flow $V^i$ at the surface for Au+Au collisions is $\sim 0.1$ at
$\tau=0.1$ fm/$c$ at midrapidity.
Fig.\ \ref{fig:7} shows the average flow fields $V^i$ for Au+Cu collisions
at impact parameters $b=0$ fm and $b=2$ fm in the $\eta-x$-plane.
In the central case the flow field leads to an expansion which is much more
pronounced on the Cu-side of the system, consistent with the rule of thumb that flux tubes
like to expand into the wake of spectators (which here are solely from the gold nucleus).
The flow pattern becomes more involved for Au+Cu collisions at finite
impact parameter.
In Fig.\ \ref{fig:8} the flow in the transverse plane is shown for forward and backward
rapidity for the $b=2$ fm Au+Cu system. We notice that the azimuthal
modulation of the flow is non-trivial  but can again be understood through
the position of spectator nucleons from the Au nucleus (centered at $x=1$ fm).
This and the previous figures make it clear that $S_-^i$ contributes not
only to directed flow but also to the elliptic flow.

The flow of energy in the classical field before $\tau_{\mathrm{th}}$ will
translate into a flow of energy in the hydrodynamic
phase after thermalization due to local energy and momentum conservation
\cite{Fries:2005yc}.
One expects remnants of this flow to survive in hydrodynamics due
to the inertia of fluid cells. In particular, this should result in a directed
flow of particles which is odd in momentum rapidity $y$. In fact such a
$y$-odd directed flow, measured by the first Fourier component $v_1$, has been
observed at RHIC \cite{Abelev:2008jga,Adams:2005ca}. The sign of the effect is consistent with the
expectation from color glass, moreover the data points as a function of
rapidity could be fitted with a $\sinh y$-shaped function. At this point it is
too early to draw strong conclusions but the coincidence of sign and shape of the
effect with data is encouraging.

Some of the qualitative features of the flow field discussed here have been
generated in hydrodynamic simulations by initializing a
tilted source \cite{Csernai:2011gg}, e.g.\ postulated in the fire streak model
\cite{Gosset:1988na}. Our calculation suggests that color glass could account for this
phenomenon without invoking additional model assumptions. In addition,
classical QCD adds several unique predictions in particular for the case of collisions
of asymmetric nuclei. A systematic study of flow as a function of rapidity and
different nuclear systems could find this unique fingerprint of color glass.

RJF acknowledges many useful discussions with J.\ Kapusta and Y.\ Li and he
is grateful for encouragement by L.\ McLerran. This project was
supported by the U.S. National Science Foundation through CAREER grant
PHY-0847538, and by the JET Collaboration and DOE grant
DE-FG02-10ER41682.

\end{document}